\documentclass[12pt]{article}

\usepackage[utf8]{inputenc}
\usepackage[english]{babel}
\usepackage{amsfonts}
\usepackage{amsmath}
\usepackage{amsthm}
 
\newtheorem{theorem}{Theorem}[section]

\newtheorem{definition}{Definition}[section]
\newtheorem{cryptosystem}{Cryptosystem}
\newtheorem*{test*}{The CPA indistinguishability experiment $\texttt{PubK}^{\texttt{cpa}}_{\mathcal{A}, \Pi}(\lambda)$} 
\newtheorem*{test1*}{The CCA indistinguishability experiment $\texttt{PubK}^{\texttt{cca}}_{\mathcal{A}, \Pi}(\lambda)$} 

\begin{document}
\begin{titlepage}
	\centering
	Summer Research Project 2016\\
	\vspace{4cm}
	{\LARGE \textbf{On the Semantic Security of NTRU}} \\
	{with a gentle introduction to cryptography}  \\
	\vspace{5cm}
	Liam Peet-Pare   \\
	\vspace{0.5cm}
	supervised by   \\
	Dr. Anne Broadbent and Dr. Raza Kazmi \\
	\vfill
	Department of Mathematics and Statistics \\
	Faculty of Science  \\
	University of Ottawa
\end{titlepage}
\pagenumbering{roman}

\section*{Abstract}
\addcontentsline{toc}{section}{Abstract}
\vspace{2cm}

This paper provides an explanation of NTRU, a post quantum encryption scheme, while also providing a gentle introduction to cryptography. NTRU is a very efficient lattice based cryptosystem that appears to be safe against attacks by quantum computers. NTRU's efficiency suggests that it is a strong candidate as an alternative to RSA, ElGamal, and ECC for the post quantum world. The paper begins with an introduction to cryptography and security proofs for cryptographic schemes before explaining the NTRU cryptosystem and culminating with a proof that the original presentation of NTRU is not IND-CPA secure. We will conclude by mentioning padding schemes to NTRU that are provably IND-CCA2 secure in the random oracle model. The paper is designed to be accessible to anyone with minimal background in abstract algebra and number theory - no previous knowledge of cryptography is assumed. Given the author's lack of familiarity with the subject, this paper aims to be an expository work rather than to provide new insights to the subject matter.
\newpage

\tableofcontents
\newpage

\pagenumbering{arabic}

\section{Introduction}

\subsection{What is Cryptography?}

Cryptography, as defined by the Oxford English Dictionary, is ``the art of writing and solving codes”. As Katz and Lindell note in \cite{modern} this is historically accurate, but with advent of rigorous modern cryptography in the 1970s and 1980s, cryptography is more a science than an art today. Cryptography refers to the twin branches of cryptology and cryptanalysis. Cryptology is the process of encrypting information, whereas cryptanalysis refers to attempts to decrypt information. Cryptography is a very old idea and has existed in various forms for thousands of years. At its core, cryptography is the art or science of sending secret messages between two parties, so that a third party cannot decipher the message even if they were to intercept the message. In \cite{stinson}, Douglas Stinson writes ``The fundamental objective of cryptography is to enable two people, usually referred to as Alice and Bob, to communicate over an insecure channel in such a way that an opponent, Oscar, cannot understand what is being said." 

In the past, cryptography was essential for sending secret messages containing sensitive information. For example, during times of war, cryptography was used to send sensitive information that could potentially be intercepted by the enemy. A famous historical example of encryption in use is the German Enigma machine during the Second World War, whose cipher was eventually broken by Alan Turing and his colleagues at Bletchley Park. In the modern digital and internet age, cryptography has become more important than ever before, as we are constantly sending sensitive information over insecure channels.

On a daily basis, most of the world relies on the information security cryptography provides without even being aware of it. For example, online banking, emails, and credit card purchases must all be encrypted to prevent third parties from stealing one's information. This increase in the scope of cryptographic applications motivates the general definition of cryptography given in \cite{modern}: ``[Cryptography is] the scientific study of techniques for securing digital information, transactions, and distributed computation.”

In our scenario with Alice, Bob, and Oscar, the unencrypted message or information that Alice wants to send to Bob is referred to as \textit{plaintext}, while the encrypted message is known as \textit{ciphertext}. The plaintext is encrypted using a \textit{key}, and is sent to Bob who is able to decrypt the ciphertext as he knows the encryption key. If the ciphertext is intercepted by Oscar, he will be unable to decrypt the message as he does not know the encryption key. 

Formally we can define a cryptosystem as follows.

\begin{definition} A cryptosystem is a five-tuple $\mathcal{(P,C,K,E,D)}$, where the following conditions are satisfied:		
\begin{enumerate}
\item $\mathcal{P}$ is a finite set of possible plaintexts.
\item $\mathcal{C}$ is a finite set of possible ciphertexts.
\item $\mathcal{K}$, the keyspace, is a finite set of possible keys.
\item For each $k \in \mathcal{K}$, there is an encryption rule $e_k \in \mathcal{E}$ and a corresponding decryption rule $d_k \in \mathcal{D}$.  Each $e_k: \mathcal{P} \to \mathcal{C}$ and $d_k: \mathcal{C} \to \mathcal{P}$ are functions such that $d_k(e_k(x))=x$ for every plaintext element $x \in \mathcal{P}.$
\end{enumerate}
\end{definition}

Modern cryptography is an intersection of computer science and mathematics. One can approach cryptography from a more mathematical perspective, or a perspective informed more from computer science. As a result, there are different vocabularies and ways of presenting information, that may seem more or less intuitive depending on one's background. The above definition is drawn from a more mathematical perspective, so an alternative computer science-centric definition can also be given. The following definition is in fact the commonly accepted definition of a cryptosystem in modern cryptography, but it is perhaps less familiar to students coming from a mathematical background. 

\begin{definition} A cryptosystem is a 3-tuple of algorithms, \\ $\Pi=(\texttt{keyGen}, \texttt{Enc}, \texttt{Dec})$ with the following properties:
\begin{enumerate}
\item The key-generation algorithm $\texttt{keyGen}(1^{\lambda})$ ($\lambda$ being the security parameter) is a probabilistic algorithm that outputs a key $k \in \mathcal{K}$ chosen according to some distribution that is determined by the scheme.
\item The encryption algorithm \texttt{Enc} takes as input a key $k$ and a plaintext $m \in \mathcal{P}$ and outputs a ciphertext $c \in \mathcal{C}$. We denote the encryption of the plaintext $m$ using the key $k$ by $\texttt{Enc}_k(m)$.
\item The decryption algorithm \texttt{Dec} takes as input a key $k$ and a ciphertext $c$ and outputs a plaintext $m$. We denote the decryption of the ciphertext $c$ using the key $k$ by $\texttt{Dec}_k(c)$.
\end{enumerate}
\end{definition}

Here, $\mathcal{K}$ is the keyspace, $\mathcal{P}$ is the set of possible plaintexts, and $\mathcal{C}$ is the set of possible ciphertexts. An encryption scheme is fully defined by specifying the three algorithms (\texttt{keyGen}; \texttt{Enc}; \texttt{Dec}) and the plaintext space $\mathcal{P}$. In order for the encryption scheme to function properly, we require that for every key $k$ output by \texttt{keyGen} and every plaintext $m \in \mathcal{P}$, it holds that $$\texttt{Dec}_k(\texttt{Enc}_k(m))=m.$$ The two definitions of a cryptosystem given above actually define a \textit{private-key} or \textit{symmetric} cryptosystem, however they can easily be modified to define a \textit{public-key} or \textit{asymmetric} cryptosystem as we will discuss later. 

While the definition of a cryptosystem is not particularly involved or complex, it does not make it clear what an encryption scheme might look like in practice, so we provide a couple of examples of simple encryption schemes borrowed from \cite{stinson} to help reify our discussion thus far.

\begin{cryptosystem} Shift Cipher \\ 
Let $\mathcal{P=K=C}=\mathbb{Z}_{26}$. For $0 \le k \le 25$, define 
$$ e_k(x)=(x+k) \ mod \ 26 $$
and
$$ d_k(y)=(y-k) \ mod \ 26 $$
$(x,y \in \mathbb{Z}_{26}).$
\end{cryptosystem}

\begin{cryptosystem} Substitution Cipher \\
Let $\mathcal{P=C}=\mathbb{Z}_{26}$. $\mathcal{K}$ consists of all possible permutations of the 26 symbols $0,1,\ldots,25$. For each permutation $\pi \in \mathcal{K}$, define
$$ e_{\pi}(x)=\pi(x), $$
and define
$$ d_{\pi}(y)=\pi^{-1}(y), $$
where $\pi^{-1}$ is the inverse permutation to $\pi$. 
\end{cryptosystem}
 
 In the Shift Cipher, our plaintext space, keyspace, and ciphertext space are all $\mathbb{Z}_{26}$. To encrypt a message we take a letter $x$ from our plaintext and add our key $k$ to it modulo 26. The resulting ciphertext has therefore been ``shifted". We provide an example to illustrate. Suppose the key for our shift cipher is $k=13$ and we have the plaintext 
 $$iamacat$$ 
 which we want to encrypt. First we convert the letters into numbers modulo 26 so our encryption key can take them as input. For our plaintext we have: 
 $$8 \ 0 \ 12 \ 0 \ 3 \ 0 \ 19$$ 
 Now we apply our encryption algorithm by adding 13 to each value and reducing each sum modulo 26. Doing this we obtain: 
 $$21 \ 13 \ 25 \ 13 \ 16 \ 13 \ 6$$ 
 To obtain our ciphertext, we now convert our numbers back to letters 
 $$vnznqng$$
Hence Alice would send ``vnznqng" to Bob. Upon receiving this ciphertext, Bob could decrypt the message to read the original ``iamacat". If Oscar intercepts the ciphertext he will not be able to decrypt it unless he knows the key, $k$. Clearly this cryptosystem is not very secure, as it would not be difficult for Oscar to determine $k$ by merely exhaustively trying all the integers $0 \le k \le 25$.  

The Shift Cipher is actually a special case of the substitution cipher, as it contains only 26 of the possible $26!$ permutations of 26 elements. The number of possible permutations, $26!$, is a very large number, so an exhaustive search for the key is infeasible. The Substitution Cipher is therefore more secure than the Shift Cipher, but there are other methods than an exhaustive search to break the cryptosystem (see \cite{stinson}) that we will not discuss here. An encryption scheme's security depends on more than an attacker being able to recover the encryption key, and we will provide a more detailed discussion of cryptographic security later in this paper. The two cryptosystems presented above can be used to convert Roman alphabet plaintexts into ciphertexts. Clearly modern day cryptography encompasses more than just sending secret messages, and the cryptosystems that are widely used today are more complex than these two toy examples. That being said, shift and substitution ciphers have been used many times throughout history, and before the invention of computers, cryptanalysing a well implemented substitution cipher would have been no trivial task.

\subsection{Public-Key Cryptography}

As we mentioned earlier, the theory of cryptography took a decisive turn in the 1970s and 1980s. The theoretical revolution in cryptography was largely sparked by a 1976 paper by Diffie and Hellman titled ``New Directions in Cryptography." This paper introduced several new ideas that would become central to the study of cryptography for the next several decades. For the purposes of this paper, the most important concept Diffie and Hellman introduced was that of \textit{public-key} or \textit{asymmetric} cryptography.

In our earlier definitions and examples of encryption schemes, both Alice and Bob shared the same key that could be used to both encrypt and decrypt messages. This is certainly an effective method of designing a cryptosystem, but its security requires that the key be kept secret from hostile third parties such as Oscar. As Alice and Bob both use the same key, and this key must be kept secret, this kind of cryptosystem is called a private-key or symmetric encryption scheme. A private-key cryptosystem requires that Alice and Bob have a secure channel in which they can share the key before they start communicating over an insecure channel. For many cryptographic purposes a private-key scheme poses no problems. For example, for military purposes it is generally feasible for personnel to physically travel to another location to share an encryption key before communicating over insecure channels, or if an individual wants to encrypt their hard-drive she merely has to ``share" the encryption key with herself.

For many other applications of cryptography however, it is clear that a private-key encryption scheme is not sufficient, as it may not be possible to exchange keys over a secure channel before commencing encrypted communication. For example, if two people who have never met and live in different countries want to communicate via encrypted email, it is less feasible to find a secure channel in which to exchange keys. Another example where a private-key scheme is even more inadequate is that of an online merchant. An online retailer has to complete thousands of secure transactions in an efficient manner with people it will never have physical contact with, so there is clearly no way to generate and securely communicate keys with all of its patrons \cite{modern}.

Thankfully, Diffie and Hellman's revolutionary idea of public-key cryptography addresses these issues. In a public-key encryption scheme, Alice and Bob have two keys - a public key and a private (or secret) key. The public key does not need to be kept secret, and it is used to encrypt messages, while the private key (which is kept secret as its name suggests) is used to decrypt messages. Under this type of encryption scheme, Bob can publish his public key (in a directory, for example), which Alice can use to encrypt her plaintext to send to Bob, and only Bob's private key will be able to decrypt the message. Once Alice encrypts her plaintext using Bob's public key, not even she will be able to decrypt the message, as Bob's secret key corresponds to his public key in some way. Note that unlike a private-key encryption scheme, a different key is used to encrypt and decrypt messages, hence public-key cryptosystems are also referred to as \textit{asymmetric} cryptosystems. To make this discussion more exact we provide a definition of a public-key cryptosystem as given by Nguyen and Pointcheval in \cite{padding}.   

\begin{definition} A public-key (or asymmetric) encryption scheme is a 3-tuple of algorithms $\Pi=(\texttt{keyGen}, \texttt{Enc}_{pk}$, $\texttt{Dec}_{sk})$ with the following properties:
\begin{enumerate}
\item The key-generation algorithm $\texttt{keyGen}(1^\lambda)$ ($\lambda$ being the security parameter), produces a pair (\texttt{pk,sk}) of public and private (secret) keys.
\item The encryption algorithm $\texttt{Enc}_{pk}(m)$ outputs a ciphertext $c$ corresponding to the plaintext $m \in \mathcal{P}$, according to the public key \texttt{pk}.
\item The decryption algorithm $\texttt{Dec}_{sk}(c)$ which outputs the plaintext $m$ associated to the ciphertext $c$ (or a distinguished failure symbol $\perp$, if $c$ is an invalid ciphertext), given the private key \texttt{sk}. 
\end{enumerate}
\end{definition}

It at first seems incredible that publishing the public key would not compromise the security of the encryption scheme, but when we look at real world analogies we can get an idea of how this might be possible. For example, everyone knows how to lock a padlock, but only a person with the key can open it once it is locked, and it is easy to break a window, but much more challenging to put it back together. Diffie and Hellman introduced the idea of public-key cryptography in 1976, but did not present a construction for an actual cryptosystem. Only one year later however, Rivest, Shamir, and Adleman introduced the RSA problem and presented a public-key encryption scheme based on it.

The RSA public-key encryption scheme is based on the hardness of factoring integers. That is, given a composite integer $N$, find positive integers $p,q$ such that $pq=N$. The general idea behind RSA is that multiplying two large primes $p$ and $q$ together is very easy, but given their product $pq=N$, factoring $N$ is extremely difficult. After the creation of the \textit{RSA Cryptosystem}, other public-key cryptosystems followed. The most important of these public-key cryptosystems are RSA and its variations and the \textit{ElGamal Cryptosystem} and its variations. As we just mentioned, the security of RSA is based on the computational infeasibility of factoring large integers, while ElGamal is based on the computational infeasibility of what is known as the discrete logarithm problem. A variation on ElGamal is known as elliptic curve cryptography (ECC) which is based on the algebraic structure of elliptic curves over finite fields. 

We will not provide details in this paper on the functioning of these cryptosystems, nor will we provide details for why the discrete logarithm problem and factoring are considered ``hard" or what exactly is meant by this term. Introductory treatments of this material can be found in \cite{modern}, \cite{math}, and \cite{stinson}. Since their inception, and especially after the advent of the internet, RSA, ElGamal, and their variations have become integral to communication all over the world.

\subsection{Post Quantum Cryptography} 

Public-key encryption schemes such as RSA, ElGamal, and ECC have proved to be extremely effective and are used all over the world to ensure secure communication in situations where private-key cryptosystems are not sufficient. As mentioned before, RSA's security is based on the hardness of factoring integers while ElGamal and ECC are based on elliptic curves, and their security depends on the hardness of the discrete logarithm problem. Thus far, no one has found a classical polynomial time algorithm to factor integers or solve the discrete logarithm problem, and it is conjectured that no such algorithm exists \cite{modern}. This is not the case however in quantum theory. 

In 1994 Peter Shor discovered quantum algorithms for factoring and discrete logarithms, now known as Shor's algorithm. Shor's algorithm can factor integers in polynomial time and it was further shown that Shor's algorithm could be applied to discrete logarithm problems and solve those in polynomial time as well. This means that as soon as a large quantum computer is developed, RSA, ElGamal, and related cryptosystems will no longer be secure, as they can be trivially broken using Shor's algorithm. No such quantum computer yet exists, but it is believed that a functional large quantum computer could be built within 20 years \cite{post}.  

While the discovery of Shor's algorithm was certainly a blow to public-key cryptography, it is far from its death knell. Despite the fact that RSA and ElGamal will no longer be secure after a large quantum computer is built, there are many other cryptographic schemes that are believed to be secure against quantum attacks. Post Quantum Cryptography therefore, refers to the study of encryption schemes that are secure against attacks by quantum computers. Some post-quantum schemes are themselves quantum (i.e. utilize the principles of quantum mechanics), while others are classical schemes. Among the potential post-quantum encryption schemes is a cryptosystem called \textit{NTRU}, which is an example of what is known as lattice based cryptography. The majority of the remainder of this paper will be devoted to explaining NTRU and exploring its security. Before we begin this discussion however, we will first need to explain some notions of \textit{security} in the world of theoretical cryptography.

\section{Security}

\subsection{What Does it Mean for an Encryption Scheme to be Secure?} 

At first glance it may appear that the notion of security for an encryption scheme is clear - if an adversary cannot break the encryption scheme then it is secure. This, however, merely begs the question of what it means to break an encryption scheme. There are several answers one might give to this question, and it immediately becomes clear that a definition of security is not so obvious after all. What, then, does it mean for a cryptosystem to be secure? There are many possible answers to this question, and we will evaluate some of them now. The following discussion is indebted to \cite{modern}.

\begin{enumerate}

\item \textit{If no adversary can find the secret key when given a ciphertext the cryptosystem is secure}. Certainly if an adversary can find the secret key the cryptosytem will be completely broken, but what if an adversary can recover the plaintext of every single ciphertext sent, but cannot actually recover the private key? Clearly the cryptosystem would not be secure. As a contrived example, consider some scheme where a message is permuted many times, but eventually returns to the original message before being sent as ciphertext. If the private key inverts this series of permutations to recover the plaintext it may be very difficult for an adversary to recover the private key, but the plaintext and ciphertext are identical, so an eavesdropper would always know the plaintext of any given ciphertext. Clearly this cryptosystem cannot be secure.

\item \textit{If no adversary can recover the plaintext that corresponds to the ciphertext, the cryptosystem is secure}. This answer is an improvement over the first answer given, but what if an adversary can recover 90\% of a plaintext, or even 99\% of a plaintext? If this is the case, then the crytposystem is also clearly not secure.

\item \textit{If no adversary can determine any meaningful information about the plaintext from the ciphertext, the cryptosystem is secure}. This answer is yet another improvement, but it is not clear what is meant by ``meaningful information." We will need to refine this definition slightly to achieve the precision we are looking for.

\item \textit{If no adversary can compute any function of the plaintext from the ciphertext, a cryptosystem is secure}. This makes our previous answer more precise, and provides very strong criteria for security. We will still have to reformulate in a way that allows us to evaluate the security of actual cryptosystems, but this is in fact what is considered to be the ``right" definition of security in modern cryptography.

\end{enumerate}

The fourth answer provided above moves us closer to a definition of security, but we still need to formalize it mathematically. Additionally, we need to consider what we assume the \textit{power} of an adversary is (e.g. computing power), and also the type of attack an adversary will conduct against an encryption scheme. 

We will first consider the power of an adversary. We will want to ensure security against any efficient adversary (i.e. any adversary running in polynomial time). We define an efficient adversary as any adversary that is capable of performing efficient computations. We refer to \cite{modern} for their definition of an efficient computation.  \\

\begin{small}``We have defined efficient computation as that which can be carried out in \textit{probabilistic polynomial time} (sometimes abbreviated \texttt{PPT}). An algorithm $A$ is said to run in polynomial time if there exists a polynomial $p(\cdot)$ such that, for every input $x \in \{0,1\}^*$, the computation of $A(x)$ terminates within at most $p(\|x\|)$ steps (here, $\|x\|$ denotes the length of the string $x$). A probabilistic algorithm is one that has the capability of ``tossing coins"; this is a metaphorical way of saying that the algorithm has access to a source of randomness that yields unbiased random bits that are each independently equal to $1$ with probability $\frac{1}{2}$ and 0 with probability $\frac{1}{2}$." \end{small} \\

It is important to define the power of an adversary because a cryptosystem's security will depend on the capability of any adversary. For example, security against an adversary running in exponential time is completely different than security against an adversary running in polynomial time. Note that while we choose to consider an efficient adversary, a different level of power could be chosen instead. We will not elaborate further on the power of an adversary, but if the reader desires, a detailed treatment can be found in \cite{modern}.

\subsection{Types of Attack} 
 
We now have to consider the \textit{type} of attack an adversary can mount. By this we do not mean the strategy that an adversary will use, as we do not want to make any assumptions about attack strategies. Making assumptions regarding the strategy that an adversary will employ is dangerous because it is always possible that an adversary could discover a novel way of attacking a cryptosystem that the cryptosystem's creators did not foresee. If we assumed that only certain strategies would be employed in attacking a cryptosystem, we could never show that an encryption scheme was safe against an unaccounted for strategy. Instead, when we refer to the type of attack an adversary can mount we mean what kind of information does an adversary have access to in order to mount their attack. In other words, we are specifying an attack scenario or model, rather than an attack strategy. We now list the main types of attack scenarios, referring to \cite{modern} for our discussion:

\begin{itemize}

\item \texttt{Ciphertext-only attack:} This is the most basic of the attack scenarios. In a ciphertext-only attack an adversary just observes a ciphertext and attempts to determine the plaintext that was encrypted.

\item \texttt{Known-plaintext attack:} In this scenario an adversary knows one or more pairs of plaintexts/ciphertexts all encrypted using the same key. The adversary then attempts to determine the plaintext corresponding to some other ciphertext for which it does not know the corresponding plaintext. 

\item \texttt{Chosen-plaintext attack:} Here, an adversary has the ability to obtain the encryption of any plaintext that it chooses. The goal of the adversary is to determine the plaintext that corresponds to some other ciphertext. Note that in a public-key encryption scheme an adversary always has the ability to mount a chosen-plaintext attack, as it has access to the public encryption key.

\item \texttt{Chosen-ciphertext attack:} This is the strongest attack scenario. In a chosen-ciphertext-attack, an adversary has access to a decryption oracle by which it can obtain the decryption of any ciphertext of its choice. Here, an adversary attempts to determine the plaintext corresponding to some other ciphertext for which it cannot obtain the decryption.

\end{itemize} 

It is important to note that these types of attack increase in their severity, and therefore form a hierarchy of security. Encryption schemes that are secure against chosen-ciphertext attacks will be secure against all other types of attacks, but the converse is not true. For example, an encryption scheme could be secure against a chosen-plaintext attack (and hence secure against the first two attack models as well), but be trivially broken by a chosen-ciphertext attack. While desirable, it is not always necessary to attain the highest degree of security. It may be appropriate to consider different attack models in different situations. For example, as mentioned above, a public-key encryption scheme ought to strive for at least security against chosen-plaintext attacks as an adversary does in fact have the ability to to obtain encryptions of plaintexts of its choice. A secret key scheme on the other hand, may not necessarily need to achieve chosen-plaintext security. 

Our preceding discussion finally enables us to state a general definition of security for a cryptographic scheme.

\begin{definition}
A cryptographic scheme for a given task is secure if no efficient adversary can compute any function of the plaintext given a ciphertext, under a certain attack model.
\end{definition}

This definition of security is a good start, but it is still too general to really make it all that useful. We will still have to elaborate further in order to come up with a truly testable notion of security.

\subsection{CPA and CCA Security}

As we mentioned in our discussion of the types of attack that an adversary can mount, a public key encryption scheme should be secure against at least chosen-plaintext attacks, as an adversary will certainly have the ability to encrypt messages using the public key. Because of this, security against chosen-plaintext attacks is in some sense the minimum required security for a public-key encryption scheme. Security against chosen-plaintext attacks is known as \textit{semantic security under CPA} or simply as \textit{CPA security}, while security against chosen-ciphertext attacks is known as \textit{CCA security}. In this section we will explain how one can test a cryptosystem to determine whether or not it is CPA secure or CCA secure. 

We will begin with a discussion of semantic security or CPA security. Semantic security more or less refers to the definition of an encryption scheme's security that we presented earlier. As we saw, however, this definition provided no straightforward way to prove the security of cryptosystems used in practice. The notion of semantic security was first proposed in 1982 by Goldwasser and Micali \cite{semantic}, but the pair later showed that this notion of security is equivalent to another definition of security known as \textit{ciphertext indistinguishability under chosen-plaintext attack} or IND-CPA. This latter notion of security finally provides us with a testable definition of security, as we will explain now. 

The IND-CPA experiment is designed to determine whether or not an adversary can distinguish between ciphertexts given some plaintexts. In the experiment, an adversary is given the public key $\texttt{pk}$ and oracle access (this means that the adversary can ask for the encryption of plaintexts and will receive ciphertexts back) to the encryption algorithm $\texttt{Enc}_{pk}$. The adversary then chooses two plaintexts $m_0$ and $m_1$, and one of the two plaintexts is randomly chosen, encrypted, and the ciphertext is sent back to the adversary. If the adversary can determine which plaintext was encrypted, they win the experiment. At first glance this may sound rather odd, so we will summarize the IND-CPA experiment below.

Let $\Pi=(\texttt{keyGen}, \texttt{Enc}_{pk}$, $\texttt{Dec}_{sk})$ be a public key cryptosystem and $\mathcal{A}$ be an efficient adversary. The CPA indistinguishability experiment is as follows: \\

\begin{test*}\textbf{:} \\
\begin{enumerate}
	\item $\texttt{keyGen}(1^\lambda)$ is run by the challenger to obtain keys $\texttt{(pk, sk)}$.
	\item Adversary $\mathcal{A}$ is given $\texttt{pk}$ as well as oracle access to $\texttt{Enc}_{pk}(\cdot)$. The adversary outputs a pair of messages $m_0, m_1 \in \mathcal{P}$ such that $m_0 \neq m_1$.  
	\item A random bit $b \in \{0,1\}$ is chosen, and the ciphertext $c \leftarrow \texttt{Enc}_{pk}(m_b)$ is computed and given to $\mathcal{A}$. The ciphertext $c$ is called the \texttt{challenge ciphertext}. At this point $\mathcal{A}$ still has access to $\texttt{Enc}_{pk}(\cdot)$.
	\item $\mathcal{A}$ outputs a bit $b' \in \{0,1\}$ which is given to the challenger. Note that $\mathcal{A}$ can compute any function that it likes on $c$.
	\item The output of the experiment is defined to be $1$ if $b'=b$, and $0$ otherwise. \\
\end{enumerate}
\end{test*} 

\begin{definition} A public-key encryption scheme $\Pi=(\texttt{keyGen}, \texttt{Enc}_{pk}$, $\texttt{Dec}_{sk})$ has indistinguishable encryptions under chosen-plaintext attacks (or is CPA secure), if for all probabilistic, polynomial time adversaries $\mathcal{A}$, there exists a negligible function \texttt{negl} such that:
$$Pr[\texttt{PubK}^{\texttt{cpa}}_{\mathcal{A}, \Pi}(\lambda)=1] \le \frac{1}{2} + \texttt{negl}(\lambda).$$
\end{definition}

The CPA indistinguishability experiment given above is just a more exact formulation of the preceding discussion, but Definition 3.2 may not be so obvious at first glance. This definition however, finally provides us with a testable notion of security. It says that a public-key cryptosystem $\Pi=(\texttt{keyGen}, \texttt{Enc}_{pk}$, $\texttt{Dec}_{sk})$ is CPA secure if an adversary's probability of success under the CPA indistinguishability experiment is less than or equal to $\frac{1}{2}$ plus some negligible function of the security parameter $\lambda$. A negligible function, \texttt{negl}, refers to a function with the property that $\texttt{negl}(\lambda) < \frac{1}{poly(\lambda)}$ for all positive $poly(\lambda)$, where $poly(\lambda)$ is a polynomial in $\lambda$.   

In other words, a cryptosystem is CPA secure if any adversary cannot win the CPA indistinguishability experiment with probability significantly more than one-half -- which would be the probability of success if the adversary were to randomly choose the bit $b'$. CPA security is considered essential for any public-key cryptosystem, but surprisingly, the original formulation of the most famous public-key cryptosystem, RSA, is not CPA secure. This is because what is now known as \textit{textbook RSA} is a deterministic rather than probabilistic cryptosystem. We will not go into detail here, but it is a nice, and in fact simple, exercise for the reader to verify that textbook RSA is not CPA secure. 

We now have a working definition to determine whether or not an encryption scheme is secure against chosen-plaintext attacks, but this tells us nothing about whether or not an encryption scheme is secure against chosen-ciphertext attacks. Allowing an adversary to have access to a decryption oracle could potentially enable it to break encryption schemes that are CPA secure, so we will need a new test for CCA security. It may seem unrealistic to allow an adversary access to a decryption oracle (i.e. allow $\mathcal{A}$ to ask for decryptions of ciphertexts of its choice), but there are many scenarios where this is a genuine risk. The reader can refer to \cite{modern} for examples of hypothetical scenarios where CCA security would be desired.

The definition of CCA security is analogous to that of CPA security, except it has the additional feature that the adversary now has access to a decryption oracle. We explain the CCA indistinguishability test and define CCA security now. \\

\begin{test1*}\textbf{:} \\
\begin{enumerate}
	\item $\texttt{keyGen}(1^\lambda)$ is run by the challenger to obtain keys $\texttt{(pk, sk)}$.
	\item Adversary $\mathcal{A}$ is given $\texttt{pk}$, oracle access to $\texttt{Enc}_{pk}(\cdot)$, and access to a decryption oracle $\texttt{Dec}_{sk}(\cdot)$. The adversary outputs a pair of messages $m_0, m_1 \in \mathcal{P}$ such that $m_0 \neq m_1$.  
	\item A random bit $b \in \{0,1\}$ is chosen, and the ciphertext $c \leftarrow \texttt{Enc}_{pk}(m_b)$ is computed and given to $\mathcal{A}$. The ciphertext $c$ is called the \texttt{challenge ciphertext}. 	
	\item $\mathcal{A}$ still has access to $\texttt{Enc}_{pk}(\cdot)$ and the decryption oracle $\texttt{Dec}_{sk}(\cdot)$, with the restriction that $\mathcal{A}$ cannot request the decryption of $c$ itself. Finally, $\mathcal{A}$ outputs a bit $b' \in \{0,1\}$ which is given to the challenger. Note that $\mathcal{A}$ can compute any function that it likes on $c$.
	\item The output of the experiment is defined to be $1$ if $b'=b$, and $0$ otherwise. \\
\end{enumerate}
\end{test1*}

\begin{definition} A public-key encryption scheme $\Pi=(\texttt{keyGen}, \texttt{Enc}_{pk}$, $\texttt{Dec}_{sk})$ has indistinguishable encryptions under chosen-ciphertext attacks (or is CCA secure), if for all probabilistic, polynomial time adversaries $\mathcal{A}$, there exists a negligible function \texttt{negl} such that:
$$Pr[\texttt{PubK}^{\texttt{cca}}_{\mathcal{A}, \Pi}(\lambda)=1] \le \frac{1}{2} + \texttt{negl}(\lambda).$$
\end{definition}

There are actually two versions of the CCA indistinguishability experiment. The first is known as CCA1 or a \textit{lunchtime attack} while the second is known as CCA2 or an \textit{adaptive chosen-ciphertext attack}. In a CCA1 attack, the adversary only has access to the decryption oracle $\texttt{Dec}_{sk}(\cdot)$ before it receives the challenge ciphertext, while a CCA2 attack allows the adversary to have access to the decryption oracle after receiving the challenge ciphertext. The CCA indistinguishability experiment we have provided is actually meant for CCA2 attacks, as the adversary $\mathcal{A}$ continues to have access to the decryption oracle even after receiving the challenge ciphertext $c$. Note that it is possible for an encryption scheme to be CCA1 secure, but not CCA2 secure. CCA2 security is a very high security requirement, and creators of encryption schemes will ideally strive to achieve this level of security, although it is very difficult to design an efficient encryption scheme that actually attains security against chosen-ciphertext attacks. 

The preceding discussion on CPA and CCA security draws on the excellent expository treatment by \cite{modern}, and the reader should refer to this work if more detail is desired.

\subsection{The Role of Assumptions}

We now have working definitions that allow us to test a cryptosystem's security under different types of attacks, but we are still missing a key ingredient for furnishing a proof of cryptographic security. The security of most encryption schemes, and in particular public-key schemes, relies on some kind of operation that would allow an adversary to break the scheme being difficult to perform. For example, as we have seen with RSA, it is easy to multiply to large prime numbers together but computationally very difficult to factor a large integer. The difficulty, however, is that integer factorization (in the worst case) is \textit{believed} to be a very hard problem, but it has not been proven that no classical polynomial time algorithm exists for integer factorization. This is not to say that there are not good reasons for believing that integer factorization is a very hard problem, but the reality is that it is still \textit{assumed} to be hard, rather than proven to be hard. An interesting side note here is that the security of RSA does not actually rely on the assumed hardness of integer factorization, but instead on the related \textit{RSA Problem}, which has not been shown to be equivalent to integer factorization \cite{rsa}. 

RSA is not unique in its reliance on an unproven assumption, and in fact, most cryptographic constructions rely on questions in computational complexity theory that are as yet unproven \cite{modern}. As a result, the security of encryption schemes often hinges on an assumption. At first blush, the central role of assumptions may make it appear that any hope of rigorous proofs of security within cryptography must be abandoned, but this is not the case. There is often good reason to believe that assumptions regarding computational complexity are true, even if we cannot prove them; so although we may not be able to provide unconditional proofs of security for many encryption schemes, a rigorous proof that can demonstrate that an encryption scheme is secure if a certain assumption is true can still be a very satisfying proof. 

Because our proofs of an encryption scheme's security will often rely on an assumption, it becomes very important to precisely and clearly state this assumption. If we are to have confidence in a security proof, we will need to know exactly what is being assumed. Proofs of security in cryptography therefore often use what can be called a \textit{reductionist approach} \cite{modern}. The reductionist approach attempts to reduce the problem given by some assumption to the problem of breaking the cryptosystem. Again, we can consider RSA as an example. RSA attempts to reduce the known hard problem of integer factorization to the problem of breaking RSA. Hence, in an ideal scenario we could prove that breaking RSA is equivalent to integer factorization (or the RSA Problem) which would give us confidence in RSA's security, as we have good reason to believe that integer factorization is a very hard problem. This is the methodology that is used in general for modern cryptography: attempt to reduce a known and well studied hard mathematical problem to the problem of breaking a cryptosystem.    

\subsection{The Random Oracle Model}

We have worked hard so far to determine how one can prove the security of a given cryptosystem, but we have not considered the question of whether or not practical encryption schemes can actually be proven secure using this methodology. The answer to this query is both yes and no. The definitions we have presented so far have been enormously successful in putting cryptography on a solid footing and ensuring that we can rigorously address the question of what it means for an encryption scheme to be secure. Thanks to this work, many provably secure cryptosystems now exist. There has, however, been quite a bit of difficulty developing provably secure \textit{efficient} encryption schemes. In practice, efficiency is almost as important as security for cryptosystems, as there is a desire and expectation that transactions be carried out quickly and efficiently, regardless of whether or not the transaction needs to be encrypted. As a result, many people would prefer to have compromised security rather than use a secure, but inefficient scheme \cite{modern}. For example, textbook RSA is not even CPA secure, and no provably secure encryption scheme based on RSA with efficiency that is comparable to textbook RSA is currently known.  

Thus far we have emphasized the importance of having provable security, but if major cryptosystems that are used in practice are not provably secure, what is the point of these rigorous notions of security? Furthermore, if we cannot prove the security of most of our efficient cryptosystems, how do we decide which encryption scheme is better to use? Attempts to address these questions have given rise to what is known as \textit{the random oracle model}. The random oracle model can be seen as a compromise between a fully rigorous proof of security and no proof whatsoever. The idea of the random oracle model is to introduce an idealized model in which we can furnish a proof of security for an encryption scheme. Even if this idealized model does not hold true in reality, the hope is that if an encryption scheme is secure in the random oracle model it will still be secure in practice, and at least does not have any fundamental security flaws \cite{modern}. The random oracle model may seem slightly strange for those who have not encountered similar ideas in the past. Before we can explain the random oracle model, however, we will need to introduce the notion of \textit{hash functions}.
 
The basic idea of a hash function is that it is a function that can take arbitrary length strings as input and compress these into shorter strings of a fixed length. Hash functions were originally used in data structures to reduce the lookup time for retrieving an element in the data structure \cite{modern}. Cryptographic hash functions have several properties that are required in order to make them useful. One of these properties is known as \textit{collision resistance}. A collision for a function $H$ is a pair of distinct inputs $x, x'$ where $x \neq x'$ such that $H(x)=H(x')$, i.e., two inputs have the same image. A function is considered collision resistant if it is infeasible for any PPT algorithm to find a collision in $H$ \cite{modern}. Other useful properties for hash functions are \textit{second preimage resistance} and \textit{preimage resistance}, but we will not go into details here. Hash functions should also be quick to compute.

The idealized situation considered in the random oracle model assumes the existence of a public, randomly chosen function $H$. This function $H$ can be evaluated only by querying an oracle. In other words, if one has an input $x$ the oracle will return $H(x)$ when queried, but no one knows how this random function $H$ actually produces this output. The function $H$ is chosen uniformly at random, which is impossible to do in polynomial time. It is perhaps easiest to think of the random oracle as a black box that parties can query and will receive a response, but have no idea what the inner workings of this black box. As we cannot actually have this random oracle in practice, we \textit{instantiate} the random function $H$ with a cryptographic has function. The proof technique for the random oracle model then comes down to proving security in a model where we assume the existence of this random oracle $H$ and then instantiating $H$ with a cryptographic hash function $\hat{H}$. That is, every time that a party would query the oracle for the value $H(x)$, the party instead computes $\hat{H}(x)$ on its own. The cryptographic has function $\hat{H}$ is in some sense providing a best \textit{approximation} of the random oracle $H$ as possible. We will not provide further technical details here, as it will not be essential to the remainder of this paper, but the interested reader can find more information on the random oracle model in \cite{modern}.   

It is worth noting that while the random oracle model does not allow one to conclusively prove the security of a cryptosystem, ``there have been no real-world attacks on any ``natural" schemes proven secure in the random oracle model" \cite{modern}. It is also true, however, that it is possible to develop contrived examples of encryption schemes which are provably secure in the random oracle model, but are insecure no matter how the random oracle is instantiated. In these contrived schemes, not only are \textit{some} instantiations of the random oracle by a cryptographic hash function insecure, but rather \textit{all} instantiations are insecure. Due to the lack of theoretical justification for the security of encryption schemes proved in this manner, the random oracle remains a controversial topic among cryptographers \cite{modern}.

\section{NTRU}

\subsection{What is NTRU?} 

The NTRU cryptosystem was first presented in 1996 at a crypto rump session by Hoffstein, Pipher, and Silverman, and first described in a publication \cite{NTRU} in 1998. NTRU is one of the fastest public key encryption schemes and it also appears to be secure against attacks by quantum computers \cite{padding}. NTRU has undergone many changes in response to security weaknesses discovered by researchers throughout the years, but the essence of its original framework has been retained. We will present the NTRU cryptosystem as described by Hoffstein, Pipher, and Silverman in \cite{math}. Before we can fully describe the NTRU cryptosystem we will need a bit of background.

\begin{definition} Fix a positive integer $N$. The ring of convolution polynomials (of rank $N$) is the quotient ring
$$ R=\frac{\mathbb{Z}[x]}{x^N-1} $$
Similarly, the ring of convolution polynomials (modulo q) is the quotient ring 
$$ R_q=\frac{(\mathbb{Z}/q\mathbb{Z})[x]}{x^N-1}$$  
\end{definition}

Note that every element in $R$ or $R_q$ has a unique representative of the form 
$$a_0 + a_1x + a_2x^2 + \cdots + a_{N-1}x^{N-1} $$
with coefficients in $\mathbb{Z}$ or $\mathbb{Z}/q\mathbb{Z}$, respectively. Operations in convolution polynomial rings are not difficult. Addition, $+$, is performed in the usual way as with polynomials (reducing coefficients mod $q$ in $R_q$) while multiplication, which we denote $\star$, is done in the usual manner as well, except we reduce mod $x^N-1$. This merely means that we require $x^N$ to be equal to $1$, so when $x^N$ appears we replace it with $1$.  

\begin{flushleft} \textbf{Notation:} A polynomial $\textbf{a}(x) \in R$ will be written as \end{flushleft}
$$ \textbf{a} = \sum_{i=0}^{N-1} a_ix^i = [a_0, a_1, \dots, a_{N-1}] $$ 

\begin{theorem}
The product of two polynomials $\textbf{a}(x),\textbf{b}(x) \in R$ is given by 
$$ \textbf{a}(x) \star \textbf{b}(x) = \textbf{c}(x)\ with\ c_k = \sum_{i+j \equiv k\ (mod\ N)} a_ib_{k-i} $$
\end{theorem}
The proof of this theorem is omitted, but it is merely an exercise in applying the definition of multiplication in $R$. Note that multiplication in $R_q$ is identical, except that coefficients will be reduced modulo $q$.

We are now ready to describe the NTRU cryptosystem. Fix an integer $N\ge1$ and two moduli $p$ and $q$, and let $R$, $R_p$, and $R_q$ be the convolution polynomial rings
$$ R=\frac{\mathbb{Z}[x]}{x^N-1}, \quad R_p=\frac{(\mathbb{Z}/p\mathbb{Z})[x]}{x^N-1}, \quad R_q=\frac{(\mathbb{Z}/q\mathbb{Z})[x]}{x^N-1}, $$
as described above. Note that in our description when we refer to computations mod $p$ we are working in $R$ mod $p$, which is equivalent to $R_p$. The same is true for $R_q$.

We require $N$ to be prime, and while $p$ and $q$ need not be prime, we require $gcd(N,q)=gcd(p,q)=1$. We will now introduce a piece of notation to make our description easier.

\begin{flushleft}
\textbf{Notation:} For any positive integers $d_1$ and $d_2$, let \end{flushleft}
\begin{displaymath}	
	\mathcal{T}(d_1,d_2) = \left\{ \textbf{a}(x) \in R:
		\begin{array}{lr}
			\textbf{a}(x)\ has\ d_1\ coefficients\ equal\ to\ 1,\\
			\textbf{a}(x)\ has\ d_2\ coefficients\ equal\ to\ -1,\\
			\textbf{a}(x)\ has\ all\ other\ coefficients\ equal\ to\ 0
		\end{array}
	\right\}
\end{displaymath}
 
Polynomials in the set $\mathcal{T}(d_1,d_2)$ are called ternary polynomials. They are similar to binary polynomials which have only 0's and 1's as coefficients. In describing NTRU we will make use of the Alice and Bob terminology introduced at the beginning of this paper. For the NTRU cryptosystem, we assign public parameters $(N,p,q,d)$ satisfying the requirements stated above. To create a private key, Alice randomly selects two polynomials 
$$ \textbf{f}(x) \in \mathcal{T}(d+1,d)  \quad and \quad \textbf{g}(x) \in \mathcal{T}(d,d). $$ 
The polynomial $\textbf{f}(x)$ must be invertible mod $p$ and mod $q$. If either inverse fails to exist, Alice discards	 this $\textbf{f}(x)$ and selects a new one. Note that $\textbf{f}(x)$ cannot be drawn from $\mathcal{T}(d,d)$ because elements of this set never have inverses (for any $\textbf{a}(x) \in \mathcal{T}(d,d),\  \textbf{a}(1)=0$). Alice next computes the inverses 
$$ \textbf{F}_p(x)=\textbf{f}(x)^{-1}\ in\ R_p \quad and \quad  \textbf{F}_q(x)=\textbf{f}(x)^{-1}\ in\ R_q. $$
Hence we have
$$ \textbf{F}_p(x) \star \textbf{f}(x) \equiv 1 \ (mod\ p) \quad and \quad \textbf{F}_q(x) \star \textbf{f}(x) \equiv 1 \ (mod\ q). $$
Alice then computes
$$ \textbf{h}(x) \equiv \textbf{F}_q(x) \star \textbf{g}(x) \quad (mod\ q). $$

Alice's public key is the polynomial $\textbf{h}(x)$ and her private key is the pair $(\textbf{f}(x),\textbf{F}_p(x))$. That is, if Bob wants to send Alice a message, he will encrypt the plaintext using $\textbf{h}(x)$ and Alice will use $(\textbf{f}(x),\textbf{F}_p(x))$ to decrypt the message. Now let's see how encryption works.

Bob's plaintext must be a polynomial $\textbf{m}(x) \in R$ with coefficients between $\frac{-1}{2}p$ and $\frac{1}{2}p$. Hence, $\textbf{m}$ is in fact the centered lift of a polynomial in $R_p$. Bob randomly chooses a polynomial $\textbf{r}(x) \in \mathcal{T}(d,d)$ and computes
$$ \textbf{e}(x) \equiv p\textbf{h}(x) \star \textbf{r}(x) + \textbf{m}(x) \quad (mod\ q). $$
 The polynomial $\textbf{e}(x)$ is Bob's ciphertext, and it is in the ring $R_q$. After computing $\textbf{e}(x)$ Bob can send this ciphertext to Alice who will decrypt it using her private key. 
 
 When Alice receives Bob's ciphertext, she begins decrypting the message by computing
 $$ \textbf{a}(x) \equiv \textbf{f}(x) \star \textbf{e}(x) \quad (mod\ q). $$
 Next, Alice center lifts $\textbf{a}$ so that its coefficients lie in the interval from $\frac{-q}{2}$ to $\frac{q}{2}$, and computes
 $$ \textbf{b}(x) \equiv \textbf{F}_p(x) \star \textbf{a}(x) \quad (mod\ p). $$

If the parameters have been chosen properly, this polynomial $\textbf{b}(x)$ is in fact equal to Bob's plaintext $\textbf{m}(x)$. In the original NTRU description, $\textbf{b}(x)$ was only equal to $\textbf{m}(x)$ with a high degree of probability, but it was later realized that this chance of decryption failure could be eliminated with a proper selection of parameters. It is not immediately clear why $\textbf{b}(x)$ is equal to $\textbf{m}(x)$, so we will explain the decryption process in more detail.

The polynomial $\textbf{a}(x)$ that Alice computes satisfies
\begin{eqnarray*}
\textbf{a}(x) &\equiv & \textbf{f}(x) \star \textbf{e}(x) \quad (mod\ q) \\
	&\equiv & \textbf{f}(x) \star (p\textbf{h}(x) \star \textbf{r}(x) + \textbf{m}(x)) \quad (mod\ q) \\
	&\equiv & \textbf{f}(x) \star (p(\textbf{F}_q(x) \star \textbf{g}(x)) \star \textbf{r}(x) + \textbf{m}(x)) \quad (mod\ q) \\    
	&\equiv & p(\textbf{f}(x) \star \textbf{F}_q(x) \star \textbf{g}(x) \star \textbf{r}(x)) + \textbf{f}(x) \star \textbf{m}(x) \quad (mod\ q) \\ 
	&\equiv & p\textbf{g}(x) \star \textbf{r}(x) + \textbf{f}(x) \star \textbf{m}(x) \quad (mod\ q)
\end{eqnarray*}

Now consider the polynomial $ p\textbf{g}(x) \star \textbf{r}(x) + \textbf{f}(x) \star \textbf{m}(x) $ computed in $R$, rather than modulo $q$. We want the polynomial to remain the same when we reduce its coefficients modulo $q$, so we need to try to bound the largest coefficient. Recall that $\textbf{g}(x)$ and $\textbf{r}(x)$ are selected from $\mathcal{T}(d,d)$, so if all of their $1$'s match up and all of their $-1$'s match up, the largest possible coefficient of $\textbf{g}(x) \star \textbf{r}(x)$ is $2d$. Similarly, we have $\textbf{f}(x) \in \mathcal{T}(d+1,d)$ and the coefficients of $\textbf{m}(x)$ lie in the interval $\frac{-1}{2}p$ to $\frac{1}{2}p$, so the largest possible coefficient of $\textbf{f}(x) \star \textbf{m}(x)$ is $(d+1)\cdot(\frac{1}{2}p) + (d)\cdot(\frac{1}{2}p) = (2d+1)\cdot(\frac{1}{2}p)$. 

Hence, if the largest coefficient of $\textbf{g}(x) \star \textbf{r}(x)$ coincides with the largest coefficient of $\textbf{f}(x) \star \textbf{m}(x)$, the largest possible coefficient for $p\textbf{g}(x) \star \textbf{r}(x) + \textbf{f}(x) \star \textbf{m}(x)$ is 
$$p\cdot2d + (2d+1)\cdot(\frac{1}{2}p) = \left(3d + \frac{1}{2}\right)p.$$
So if we take $q > (6d+1)p$ then the largest coefficient will be strictly less than $\frac{1}{2}q$. Therefore, when Alice computes $\textbf{a}(x)$ modulo $q$ and center lifts it to $R$, she in fact recovers exactly the polynomial $p\textbf{g}(x) \star \textbf{r}(x) + \textbf{f}(x) \star \textbf{m}(x)$. That is, we actually have 
$$\textbf{a}(x) = p\textbf{g}(x) \star \textbf{r}(x) + \textbf{f}(x) \star \textbf{m}(x)$$
in $R$, rather than just a congruence modulo $q$. 

Now we can finally explain the last step of decryption, in which Alice multiplies $\textbf{a}(x)$ by $\textbf{F}_p(x)$ and reduces the product modulo $p$
\begin{eqnarray*}
\textbf{b}(x) &=& \textbf{F}_p(x) \star \textbf{a}(x) \\
	&=& \textbf{F}_p(x) \star (p\textbf{g}(x) \star \textbf{r}(x) + \textbf{f}(x) \star \textbf{m}(x)) \\
	&\equiv & \textbf{F}_p(x) \star \textbf{f}(x) \star \textbf{m}(x) \quad (mod\ p) \\
	&\equiv & \textbf{m}(x) \quad (mod\ p).	
\end{eqnarray*}	
Hence, we have shown that $\textbf{b}(x)$ and $\textbf{m}(x)$ are in fact the same modulo $p$.

This leads to the following theorem:

\begin{theorem}
If the NTRU parameters $(N,p,q,d)$ are chosen to satisfy
$$ q > (6d+1)p, $$
then the polynomial $\textbf{b}(x)$ is equal to Bob's plaintext $\textbf{m}(x)$.
\end{theorem}
The proof of the theorem was presented in our explanation of the decryption process. Note that our discussion of NTRU follows that presented in \cite{math}.

From this description of NTRU, it is not immediately clear why it is referred to as a  ``lattice based cryptosystem", and we will not provide much detail on this point as it is outside of the scope of this paper. In lieu of an in depth discussion, we will provide a couple of basic definitions and recommend further reading for those who are interested. We refer to \cite{math} for our discussion of lattices. 

\begin{definition} Let $\textbf{v}_1,\dots,\textbf{v}_n \in \mathbb{R}^m$ be a set of linearly independent vectors. The lattice $\mathcal{L}$ generated by $\textbf{v}_1,\dots,\textbf{v}_n$ is the set of linear combinations of $\textbf{v}_1,\dots,\textbf{v}_n$ with coefficients in $\mathbb{Z}$,
$$\mathcal{L} = \{a_1\textbf{v}_1 + a_2\textbf{v}_2 + \cdots +a_n\textbf{v}_n : a_1, a_2, \dots, a_n \in \mathbb{Z}\}.$$
\end{definition} 

A helpful way of thinking about a lattice is as a vector space, but over $\mathbb{Z}$ instead of a field. As with vector spaces, a \textit{basis} for a lattice $\mathcal{L}$ is any set of linearly independent vectors that generates $\mathcal{L}$ while the \textit{dimension} of $\mathcal{L}$ is the number of vectors in a basis for $\mathcal{L}$. Note that one can equivalently define a lattice as a discrete additive subgroup of $\mathbb{R}^n$ \cite{survey}. 

Once we have a lattice, we can define difficult computational problems. Two of these fundamental lattice problems are: 

\begin{definition} \textbf{(Shortest Vector Problem (SVP)).} Given an arbitrary basis of some lattice $\mathcal{L}$, find a shortest nonzero lattice vector, i.e., find a $\textbf{v} \in \mathcal{L}$ such that $\|\textbf{v}\| = \min\limits_{\textbf{v} \in \mathcal{L} \setminus \{0\}} \|\textbf{v}\|$
\end{definition}

\begin{definition} \textbf{(Closest Vector Problem (CVP)).} Given a vector $\textbf{w} \in \mathbb{R}^m$ that is not in $\mathcal{L}$, find a vector $\textbf{v} \in \mathcal{L}$ that is closest to $\textbf{w}$, i.e., find a vector $\textbf{v} \in \mathcal{L}$ that minimizes $\|\textbf{w} - \textbf{v}\|$.
\end{definition}

Note that in the above definitions $\| \cdot \|$ is the Euclidean norm. Both SVP and CVP are considered to be very hard problems computationally, and as we have seen with integer factorization and RSA, it is possible to leverage the hardness of these (or related) lattice problems to create a secure cryptosystem. Without going into any details, we mention that the NTRU problem is related to SVP for key recovery and to CVP for plaintext recovery, and to another problem known as \textit{Ring Learning with Errors} \cite{math},\cite{survey}. It is important to note, however, that ``there is no known reduction from any worst-case lattice problem to any standard version of the NTRU problem, nor from the the NTRU problem to breaking the cryptosystem's semantic security" \cite{survey}. There is a rich theory and deep mathematical questions involved in lattice based cryptography, and, if the reader is interested in delving into this material, there are many readily available sources including \cite{survey}, \cite{math}, \cite{modern}, \cite{post}, and \cite{stinson}.

\subsection{The Semantic Security of NTRU}

The NTRU cryptosystem as originally presented, and also the slightly modified description we have presented, is in fact not semantically secure (i.e. CPA secure). The original presentation of NTRU in \cite{NTRU} did not make any mention of the modern notions of cryptographic security such as CPA and CCA security, so it is unclear as to whether or not the authors were aware that their construction was not CPA secure. It is surprisingly easy to demonstrate that NTRU is not CPA secure, and it is has been mentioned in several papers including \cite{padding} and in an updated description of NTRU released by NTRU Cryptosystems and including Daniel Lieman as an author. 

The failure to provide semantic security comes from the construction of the encrypted messages. Recall that an encrypted message has the form $ \textbf{e}(x) \equiv p\textbf{h}(x) \star \textbf{r}(x) + \textbf{m}(x) \quad (mod\ q). $ Now, we will also recall that $\textbf{m}(x) \in R$, $\textbf{r}(x) \in \mathcal{T}(d,d)$, and $\textbf{h}(x) \equiv \textbf{F}_q(x) \star \textbf{g}(x) \quad (mod\ q).$ Finally, we also note that $\textbf{g}(x) \in \mathcal{T}(d,d).$ Now if we remember our earlier definition, we can see that all polynomials in $\mathcal{T}(d,d)$ have $d$ coefficients equal to $1$, $d$ coefficients equal to $-1$, and all other coefficients equal to $0$. Hence, any polynomial in $\mathcal{T}(d,d)$ evaluated at $x=1$ will be equal to zero. That is, $\textbf{r}(1) = 0$ and $\textbf{g}(1) = 0$ (and hence $\textbf{h}(1) \equiv 0 \quad (mod\ q)$). Therefore if we evaluate the encrypted message $\textbf{e}(x)$ at $x=1$ we will have 

\begin{eqnarray*}
\textbf{e}(1) &\equiv & p\textbf{h}(1) \star \textbf{r}(1) + \textbf{m}(1) \quad (mod\ q) \\
	&\equiv & p0 \star 0 + \textbf{m}(1) \quad (mod\ q) \\
	&\equiv & \textbf{m}(1) \quad (mod\ q)
\end{eqnarray*}

Using this knowledge, we will now refer to our test of CPA security to show why NTRU is not semantically secure. Recall that the adversary, $\mathcal{A}$, is allowed to choose two messages $m_0, m_1 \in \mathcal{P}$ such that $m_0 \not= m_1$. In particular, the adversary is able to choose $\textbf{m}_0(x), \textbf{m}_1(x) \in \mathcal{R}$ such that $\textbf{m}_0(1) \not\equiv  \textbf{m}_1(1) \quad (mod\ q)$. Next, a random bit $b \in \{0,1\}$ is chosen and the challenge ciphertext $c \leftarrow \texttt{Enc}_{pk}(m_b)$ is computed and given to $\mathcal{A}$. In the case of NTRU we will have $c= \textbf{e}(x) \equiv p\textbf{h}(x) \star \textbf{r}(x) + \textbf{m}_b(x) \quad (mod\ q)$ for some $p$, $\textbf{h}(x)$, and $\textbf{r}(x)$ from their respective spaces. At this point the adversary can compute any function it likes on $c$, and, in particular, it can evaluate $c=\textbf{e}(x)$ at $x=1$. Now when $\mathcal{A}$ evaluates $c$ at $x=1$ it will get 
$$\textbf{e}(1) \equiv \textbf{m}_b(1) \quad (mod\ q).$$
As the adversary has chosen two messages $\textbf{m}_0, \textbf{m}_1$ such that \\ $\textbf{m}_0(1) \not\equiv \textbf{m}_1(1) \quad (mod\ q)$, it will now know precisely which message was encrypted to give the challenge ciphertext. Hence, $\mathcal{A}$ can win the CPA security experiment with probability equal to one -- so NTRU is not CPA secure.

\subsection{Padding Schemes and NTRU Today}

After it became clear that NTRU was not semantically secure, its creators introduced several padding schemes with the aim of improving NTRU's security, and, in particular, achieving semantic security. In response to these proposed padding schemes, Phong Q. Nguyen and David Pointcheval published \cite{padding} in which they explore NTRU's padding schemes and propose several of their own. Nguyen and Pointcheval carefully explain the NTRU assumption, and demonstrate that, contrary to claims from NTRU's creators, the three padding schemes are not IND-CCA2 secure, and in fact the first is not even IND-CPA secure. The authors then propose two padding schemes of their own, which are provably secure in the random oracle model. 

The two schemes which they present are based on OAEP (Optimal Asymmetric Encryption Padding) and REACT (Rapid Enhanced-security Asymmetric Cryptosystem Transform) respectively. OAEP and REACT are padding schemes that can be applied to any cryptosystem which satisfies certain properties, and are not unique to NTRU. For example, OAEP is generally used in combination with RSA \cite{padding}. We will not go into the technical details regarding these padding schemes, as an understanding of the schemes and their security requires further knowledge of cryptography, but we hope the interested reader will explore these schemes on their own.    

NTRU continues to be seen as a strong candidate for a post quantum public-key encryption scheme, but there still needs to be more research into NTRU in order to build more confidence about the cryptosystem's security. In 2011, Damien Stehle and Ron Steinfeld developed a version of NTRU that is provably secure in the standard model, but in achieving this level of security the cryptosystem became too inefficient to be practically used \cite{pntru}. Steinfeld further extended this work in \cite{strength}, but the resulting modifications to the encryption scheme are still too inefficient. Whether it is through a padding scheme, or some other alteration, we can hope that a provably secure version of NTRU, that maintains its efficiency, will be discovered sometime in the near future.


\vfill

\addcontentsline{toc}{section}{References}

\begin{thebibliography}{20}

\bibitem{post} Daniel J. Bernstein, Johannes Buchmann, and Erik Dahmen. {\em Post Quantum Cryptography}. 2009: Springer-Verlag, Berlin.

\bibitem{rsa} Dan Boneh and Ramarathnam Venkatesan. (1998). Breaking RSA may not be equivalent to factoring. {\em Advances in Cryptology - EUROCRYPT '98. Lecture Notes in Computer Science, Volume 1403}, 59-71. Springer, Berlin.

\bibitem{semantic} Shafi Goldwasser and Silvio Micali. (1982). Probabilistic Encryption \& How to Play Mental Poker Keeping Secret all Partial Information. {\em STOC '82 Proceedings of the fourteenth annual ACM symposium on Theory of computing} (pp. 365-377). ACM, New York.

\bibitem{NTRU} Jeffrey Hoffstein, Jill Pipher, and Joseph H. Silverman. (1998). NTRU: A ring based public key cryptosystem. {\em Lecture Notes in Computer Science, Volume 1423}, 267-288. Springer, Berlin.

\bibitem{math} Jeffrey Hoffstein, Jill Pipher, and Joseph H. Silverman. {\em An Introduction to Mathematical Cryptography}. 2008: Springer, N.Y.

\bibitem{modern} Jonathan Katz and Yehuda Lindell. {\em Introduction to Modern Cryptography}. 2007: CRC Press, N.Y.

\bibitem{ring} Vadim Lyubashevsky, Chris Peikert, and Oded Regev. (2010). On Ideal Lattices and Learning with Errors over Rings. In H. Gilbert (ed.), {\em Advances in Cryptology - EUROCRYPT 2010} (pp. 1-23). Springer, Berlin.

\bibitem{padding} Phong Q. Nguyen and David Pointcheval. (2002). Analysis and Improvements of NTRU Encyption Paddings. In M. Yung (ed.), {\em Advances in Cryptology - Proceedings of CRYPTO 2002} (pp. 210-225). Springer-Verlag, Berlin.

\bibitem{survey} Chis Peikert. {\em A Decade of Lattice Cryptography} 2016: Now Publishers.

\bibitem{pntru} Damien Stehle and Ron Steinfeld. (2011). Making NTRU as Secure as Worst-Case Problems over Ideal Lattices. In K.G. Patterson (ed.), {\em Advances in Cryptology - EUROCRYPT 2011} (pp. 27-47). Springer, Berlin.

\bibitem{strength} Ron Steinfeld et al. (2012). NTRUCCA: How to Strengthen NTRUEncrypt to Chosen-Ciphertext Security in the Standard Model. {\em Lecture Notes in Computer Science, Volume 7293}, 353-371. Springer, Berlin.

\bibitem{stinson} Douglas Stinson. {\em Cryptography Theory and Practice}. 2002: CRC Press, Boca Raton.



\end{thebibliography}
\end{document}